\documentclass[prd,twocolumn,nofootinbib,showpacs,showkeys]{revtex4}
\usepackage{bm,amsmath,amssymb}
\newcommand\gothfamily{\usefont{U}{ygoth}{m}{n}}
\DeclareTextFontCommand{\textgoth}{\gothfamily}

\begin{document}

\title{On the mass of the Universe born in a black hole}
\author{Nikodem J. Pop{\l}awski}
\affiliation{Department of Physics, Indiana University, Swain Hall West, 727 East Third Street, Bloomington, Indiana 47405, USA}
\email{nipoplaw@indiana.edu}
\date{\today}

\begin{abstract}
It is shown, using the Einstein-Cartan-Sciama-Kibble theory of gravity, that gravitational collapse of spin-fluid fermionic matter with a stiff equation of state in a black hole of mass $M$ forms a new universe of mass $\sim M_\ast=M^2 m_n/m_\textrm{Pl}^2$, where $m_n$ is the mass of a neutron.
Equaling $M_\ast$ to the mass of the Universe, which is about $10^{26}$ solar masses, gives $M\sim 10^3$ solar masses.
Our Universe may thus have originated from the interior of an intermediate-mass black hole.
\end{abstract}

\pacs{04.50.Kd, 98.80.Bp}
\keywords{torsion, spin fluid, stiff matter, bouncing cosmology.}
\maketitle

In a recent paper \cite{infl}, we proposed a simple scenario which explains why our Universe appears spatially flat, homogeneous and isotropic.
We used the Einstein-Cartan-Sciama-Kibble (ECSK) theory of gravity which naturally extends the general theory of relativity (GR) to account for the quantum-mechanical, intrinsic angular momentum (spin) of elementary particles that compose gravitating matter \cite{torsion,Lo}.
The ECSK theory of gravity is based on the Lagrangian density for the gravitational field that is proportional to the curvature scalar, as in GR \cite{LL}.
It removes, however, the constraint of GR that the torsion tensor (the antisymmetric part of the affine connection) be zero by promoting this tensor to a dynamical variable like the metric tensor \cite{torsion,Lo}.
The torsion tensor is then given by the principle of stationary action and in many physical situations it turns out to be zero.
But in the presence of fermions, which compose all stars in the Universe, spacetime torsion does not vanish because Dirac fields couple minimally to the torsion tensor \cite{torsion,Lo}.
At macroscopic scales, such particles can be averaged and described as a spin fluid \cite{spin_fluid}.

The Einstein-Cartan field equations of the ECSK gravity can be written as the general-relativistic Einstein equations with the modified energy-momentum tensor \cite{torsion,Lo}.
Such a tensor has terms which are quadratic in the spin tensor and thus do not vanish after averaging \cite{avert_avg}.
These terms are significant only at densities of matter that are much larger than the density of nuclear matter.
They generate gravitational repulsion in spin-fluid fermionic matter \cite{avert_avg,avert}.
Such a repulsion becomes significant at extremely high densities that existed in the early Universe and exist inside black holes.
It prevents a cosmological singularity, replacing it by a state of minimum but finite radius.
This extremely hot and dense state is a bounce that follows the contracting phase of the Universe and initiates its rapid expansion \cite{infl,bounce}.
Torsion therefore provides a natural way to explain what caused such an expansion and what existed before the Universe began to expand.

In \cite{infl}, we showed that the dynamics of the closed Universe immediately after the bounce naturally solves the flatness and horizon problems in cosmology because of an extremely small and negative torsion density parameter.
Thus the ECKS gravity not only eliminates the problem of the initial singularity but also provides a compelling alternative to cosmic inflation.
We also suggested that the coupling between spin and torsion may be the mechanism that allows for a scenario in which every black hole produces a new universe inside, instead of a singularity.
The contraction of our Universe before the bounce at the minimum radius may thus correspond to the dynamics of matter inside a collapsing black hole existing in another universe \cite{BH}.
A scenario of the Universe born in a black hole seems more reasonable than the contraction of the Universe from infinity in the past because the latter does not explain what caused such a contraction.
This scenario also explains the time asymmetry in a new universe forming inside a black hole because the motion of the collapsing matter through the black hole's event horizon can only happen in one direction and thus it is asymmetric \cite{infl}.
The arrow of time in a daughter universe would thus be inherited, through torsion, from a mother universe.

Torsion appears as a plausible physical phenomenon that may solve some major puzzles regarding our understanding of elementary particles, black holes and the Universe.
It may introduce an effective ultraviolet cutoff in quantum field theory for fermions \cite{non}.
Moreover, torsion modifies the classical Dirac equation by generating the cubic Hehl-Datta term \cite{Dirac} which may be the source of the observed matter-antimatter imbalance and dark matter in the Universe \cite{bar}.
Torsion can also generate massive vectors that are characteristic to electroweak interactions \cite{mass}.
Finally, the gravitational interaction of condensing fermions due to the Hehl-Datta term may be the source of the observed small, positive cosmological constant which is the simplest explanation for dark energy that accelerates the present Universe \cite{dark}.

Observations of thermal emission during X-ray bursts suggest that the matter inside neutron stars is characterized by a stiff equation of state $p=\epsilon$ \cite{stiff_NS}, where $p$ is the pressure and $\epsilon$ is the energy density.
Theoretical models of ultradense matter predict the stiff equation of state at very high densities from the strong interaction of the nucleon gas \cite{stiff_mat}.
The matter collapsing inside stellar black holes is thus expected to be stiff too.
Stiff matter has also been suggested as a possible content of the Universe during early stages of its expansion \cite{stiff_cosm,stiff_BH}.
Furthermore, a black hole grows as fast as the Universe if the equation of state is stiff \cite{stiff_BH}, showing that the black hole's interior composed of stiff matter is dynamically equivalent to a universe.

If the matter collapsing in a forming black hole is stiff then the covariant conservation of the modified energy-momentum tensor causes that the mass of the universe inside this black hole increases by many orders of magnitude, as we show below.
Physically, such an increase of mass is realized by particle production in the presence of strong gravitational fields \cite{part}.
Such a pair production does not change the equation of state of the collapsing matter, which remains stiff because of the ultradense regime.
It also does not change the total (matter plus gravitational field) energy of the universe in a black hole, which is equal to zero (even in the presence of a cosmological constant) \cite{energy}.
The formation and evolution of such a universe (including the above increase of its mass) is not visible for observers outside the black hole, for whom the event horizon's formation and all subsequent processes would occur after an infinite amount of time had elapsed \cite{LL}.
Gravitational time dilation thus makes it possible for the mass of the Universe to be much bigger than the mass of the parent black hole as measured by external observers.
In this paper, we derive such a mass as a function of the mass of the parent black hole.

The Einstein-Cartan field equations for the closed Friedman-Lema\^{i}tre-Robertson-Walker (FLRW) metric, describing a closed ($k=1$), homogeneous and isotropic universe, are given by the Friedman equations for the scale factor $a(t)$ \cite{infl,spin}:
\begin{eqnarray}
& & {\dot{a}}^2+k=\frac{1}{3}\kappa\Bigl(\epsilon-\frac{1}{4}\kappa s^2\Bigr)a^2+\frac{1}{3}\Lambda a^2, \label{Fri1} \\
& & \frac{d}{dt}\bigl((\epsilon-\kappa s^2/4)a^3\bigr)+(p-\kappa s^2/4)\frac{d}{dt}(a^3)=0,
\label{Fri2}
\end{eqnarray}
where dot denotes differentiation with respect to $ct$ and $s^2$ is the square of the spin density \cite{infl}.
For a fluid consisting of unpolarized fermions, this quantity is given by
\begin{equation}
s^2=\frac{1}{8}(\hbar cn)^2,
\label{spide}
\end{equation}
where $n$ is the particle number density \cite{spin}.
Equation (\ref{Fri2}) corresponds to the covariant conservation of the modified energy-momentum tensor.
For the stiff equation of state and for densities at which $\kappa s^2\ll\epsilon$, this equation gives $\epsilon\propto a^{-6}$, which then gives $E\propto a^{-3}$, where $E$ is the total energy of matter in the Universe.
Thus the total mass of matter in the Universe $m$ scales according to $m\propto a^{-3}$, which gives $N\propto a^{-3}$, where $N$ is the number of fermions in the Universe (we assume that this number is proportional to the total number of particles).
Accordingly, $n\propto a^{-6}$, from which we obtain $s^2\propto a^{-12}$ because of (\ref{spide}).\footnote{
We cannot use the formula $dn/n=d\epsilon/(\epsilon+p)$ because it is valid only if the number of particles is conserved.
In this case, $p=\epsilon$ would give $n\propto\epsilon^{1/2}\propto a^{-3}$, which yields $N\propto\mbox{const}$.
}
Substituting (\ref{spide}) into the Friedman equation (\ref{Fri1}) gives then
\begin{equation}
{\dot{a}}^2+k=\frac{1}{3}\kappa\epsilon_0\frac{a_0^6}{a^4}-\frac{1}{96}(\hbar c\kappa)^2 n_0^2\frac{a_0^{12}}{a^{10}},
\label{master}
\end{equation}
where $\epsilon_0$ and $n_0$ correspond to the universe at $a=a_0$.
We neglect the last term in (\ref{Fri1}) with the cosmological constant $\Lambda$ because it is negligibly small for $a<a_0$.

A universe born in a black hole begins to contract when $\dot{a}=0$ at $a_0=r_g$, where $r_g=2GM/c^2$ is the Schwarzschild radius of the black hole of mass $M$.
At this stage, the second term on the right-hand side of (\ref{master}) is negligible, yielding
\begin{equation}
\epsilon_0=\frac{Mc^2}{4\pi r_g^3/3},
\label{enerde}
\end{equation}
which coincides with the rest energy density of the black hole regarded as a uniform sphere.
If the matter in a black hole at $a=a_0$ is composed of neutrons, which is a reasonable assumption, then the initial number density of fermions in this universe is
\begin{equation}
n_0=\frac{3M/m_n}{4\pi r_g^3/3},
\label{nude}
\end{equation}
where $m_n$ is the mass of a neutron.
The universe in the black hole contracts until the negative second term on the right-hand side of (\ref{master}) becomes significant and counters the first term ($k$ is negligible in this regime).
At this moment, where $a=a_\textrm{min}$, the condition $\dot{a}=0$ in (\ref{master}) gives $a_\textrm{min}=(\kappa/(32\epsilon_0))^{1/6}(\hbar cn_0)^{1/3}a_0$.
Substituting (\ref{enerde}) and (\ref{nude}) into this relation gives the scale factor of the universe at the bounce, $a_\textrm{min}$:
\begin{equation}
a_\textrm{min}=\Bigl(\frac{27}{128}\Bigr)^{1/6}\frac{2G\tilde{M}}{c^2},
\label{amin}
\end{equation}
where
\begin{equation}
\tilde{M}=\Bigl(\frac{m_\textrm{Pl}^2 M^2}{m_n}\Bigr)^{1/3}.
\label{mtilde}
\end{equation}
The mass of the universe at the bounce, $m_\textrm{max}$, is given by $m_\textrm{max}=(a_0/a_\textrm{min})^3 M$ (since $m\propto a^{-3}$).
Substituting (\ref{amin}) into this relation gives
\begin{equation}
m_\textrm{max}=\Bigl(\frac{27}{128}\Bigr)^{-1/2}M_\ast,
\label{mmax}
\end{equation}
where
\begin{equation}
M_\ast=\frac{M^2 m_n}{m_\textrm{Pl}^2}.
\label{mstar}
\end{equation}
For the mass of a typical stellar black hole, $M=10\,M_\odot$, we find
\begin{equation}
a_\textrm{min}\approx 6\times 10^{-3}\,\mbox{m},\,\,\,m_\textrm{max}\approx 3.1\times 10^{51}\,\mbox{kg}.
\label{masuni}
\end{equation}

After the bounce, the matter in a black hole expands.
If the equation of state of the matter remains stiff then the expansion of the universe in such a black hole would look like the time reversal of the contraction before the bounce, ending at $a=a_0$.
The universe would then contract again to $a_\textrm{min}$.
Such a closed universe would be cyclic, oscillating between a finite minimum scale factor $a_\textrm{min}$ caused by the gravitational repulsion from the spin-torsion coupling and a finite maximum scale factor $a_0$ existing because of $k=1$.
If, however, the equation of state of the matter near the bounce becomes relativistic, which can occur as result of the annihilation (to photons) of particle-antiparticle pairs created during the contracting phase, then the universe will expand beyond $a_0$.

The particle-antiparticle annihilation lowers the rate at which the number density of fermions (and thus the spin density) increases during the contracting phase.
But a lower spin density delays reaching the regime when the gravitational repulsion due to spin/torsion becomes significant.
Thus the universe in a black hole contracts to a scale factor lower than (\ref{amin}), resulting in a further increase of the mass inside the black hole.
The net mass of the baryonic component of the universe at the bounce is on the order of that in (\ref{mmax}).
If the baryon-to-photon energy ratio is $\eta$ then the photonic component of the universe has the energy on the order of $M_\ast c^2/\eta$.
The resulting minimum scale factor would correspond to $n_0$ in (\ref{nude}) multiplied by $\eta$, which is equivalent to (\ref{amin}) with $m_n$ divided by $\eta$.
To obtain the minimum scale factor $a_\textrm{m}=9\times 10^{-6}\,\mbox{m}$, derived in \cite{infl}, we need
\begin{equation}
\eta\approx\Bigl(\frac{a_\textrm{m}}{a_\textrm{min}}\Bigr)^3\approx 3.3\times 10^{-9}.
\end{equation}

The conservation equation (\ref{Fri2}) for ultrarelativistic matter ($p=\epsilon/3$) gives $\epsilon\propto a^{-4}$ and $s^2\propto a^{-6}$ \cite{infl,spin}, which then gives $E\propto a^{-1}$.
Thus the total mass of matter in the Universe after the bounce scales according to $m\propto a^{-1}$, until the matter becomes nonrelativistic ($p\approx 0$), $\epsilon\propto a^{-3}$ and $E\propto m\propto\mbox{const}$.
The Friedman equation (\ref{Fri1}) during the expanding phase for relativistic matter with spin is
\begin{equation}
{\dot{a}}^2+k=\frac{1}{3}\kappa\epsilon_\textrm{max}\frac{a_\textrm{min}^4}{a^2}-\frac{1}{96}(\hbar c\kappa)^2 n_\textrm{max}^2\frac{a_\textrm{min}^6}{a^4},
\label{exp}
\end{equation}
where $\epsilon_\textrm{max}=(a_0/a_\textrm{min})^6\epsilon_0$ and $n_\textrm{max}=(a_0/a_\textrm{min})^6 n_0$ correspond to the universe at $a=a_\textrm{min}$.
The universe expands until $a=a_\textrm{max}$ when the first term on the right-hand side of (\ref{master}) is equal to $k$ (the second term on the right-hand side of (\ref{exp}) is negligibly small in this regime), which gives
\begin{equation}
a_\textrm{max}=\frac{a_0^2}{a_\textrm{min}}\propto M^{4/3}.
\label{amax}
\end{equation}
The mass of the universe at this moment is equal to $m_\textrm{min}=(a_\textrm{min}/a_\textrm{max}) m_\textrm{max}$,
which gives
\begin{equation}
m_\textrm{min}=m_\textrm{max}\Bigl(\frac{a_\textrm{min}}{a_0}\Bigr)^2\propto M^{4/3}.
\label{mmin}
\end{equation}
For $M=10\,M_\odot$, we find $a_\textrm{max}\approx 1.5\times 10^{11}\,\mbox{m}$ and $m_\textrm{min}\approx 1.2\times 10^{38}\,\mbox{kg}$.
After reaching $a_\textrm{max}$, the universe contracts back to $a_\textrm{min}$.
Such a universe would be also cyclic, oscillating between $a_\textrm{min}$ and $a_\textrm{max}$.
Including the cosmological-constant term $\Lambda a^2/3$ on the right-hand side of (\ref{exp}) would not change this behavior because such a term is negligibly small even at $a=a_\textrm{max}$.

Since the above $a_\textrm{max}$ is only $\approx 1\,\mbox{AU}$, the expanding phase of our Universe cannot be described by (\ref{exp}).
In the early Universe, however, fermionic matter may be dominated by extremely massive particles.
Furthermore, fermions at very high densities acquire additional effective masses due to the cubic term in the Hehl-Datta equation, which may also be responsible for baryogenesis \cite{bar}.
Thus fermions in the early Universe may be effectively nonrelativistic.
In this case, the Friedman equation (\ref{exp}) should be replaced by
\begin{eqnarray}
& & {\dot{a}}^2+k=\frac{1}{3}\kappa\epsilon_\textrm{max}^\textrm{R}\frac{a_\textrm{min}^4}{a^2}+\frac{1}{3}\kappa\epsilon_\textrm{max}^\textrm{M}\frac{a_\textrm{min}^3}{a} \nonumber \\
& & -\frac{1}{96}(\hbar c\kappa)^2 n_\textrm{max}^2\frac{a_\textrm{min}^6}{a^4}+\frac{1}{3}\Lambda a^2,
\label{nonrel}
\end{eqnarray}
where R denotes radiation, M denotes nonrelativistic matter, $\epsilon_\textrm{max}^\textrm{R}=\epsilon_\textrm{max}/\eta$, and $\epsilon_\textrm{max}^\textrm{M}=\epsilon_\textrm{max}$.
As the Universe expands, the third term on the right-hand side of (\ref{nonrel}) becomes negligibly small.
Eventually, the first term decreases below the second term and nonrelativistic matter begins to dominate over radiation.
The Friedman equation (\ref{nonrel}) in the matter-dominated Universe reduces, using $\kappa\epsilon_\textrm{max}a_\textrm{min}^3/3=2Gm_\textrm{max}/c^2$, to
\begin{equation}
{\dot{a}}^2+k=\frac{b}{a}+\frac{1}{3}\Lambda a^2,
\label{mat}
\end{equation}
where
\begin{equation}
b=\frac{2Gm_\textrm{max}}{c^2}.
\end{equation}
The mass of such a Universe is $m\approx m_\textrm{max}\sim M_\ast$.

If the cosmological constant exceeds a critical value $\Lambda_c=4/(9b^2)$ (if $b$ exceeds a critical value $b_c=2\Lambda^{-1/2}/3$) then a universe described by (\ref{mat}) expands to infinity, starting accelerating at a scale factor $a_a=(3b/(2\Lambda))^{1/3}$ \cite{Lo,dyn}.
If $\Lambda<\Lambda_c$ ($b<b_c$) then the universe reaches a maximum scale factor $a_\textrm{max}(b,\Lambda)$ and then it contracts back to $a_\textrm{min}$.
Such a universe would be cyclic, oscillating between $a_\textrm{min}$ and $a_\textrm{max}$.
If $\Lambda=\Lambda_c$ ($b=b_c$) then the universe asymptotically tends to a static state with a maximum scale factor $a_\infty=\Lambda_c^{-1/2}=3b_c/2$.
For the universe in a black hole to be able to expand to infinity, its mass $m\approx m_\textrm{max}$ must be thus larger than a critical mass $m_\textrm{c}$:
\begin{equation}
m_\textrm{max}>m_\textrm{c}=\frac{c^2}{3G\sqrt{\Lambda}}.
\label{mcri}
\end{equation}
The relations (\ref{mmax}), (\ref{mstar}) and (\ref{mcri}) give thus
\begin{equation}
M>M_c=\Bigl(\frac{128}{27\Lambda}\Bigr)^{1/4}\frac{m_\textrm{Pl}c}{(3Gm_n)^{1/2}}.
\label{mbh}
\end{equation}
Black holes with masses below the critical mass $M_c$ produce cyclic universes, whereas black holes with masses above $M_c$ produce universes that expand to infinity.

The cosmological constant is given by $\Lambda=3\Omega_\Lambda H^2/c^2$, where $\Omega_\Lambda\approx 0.74$ is the vacuum density parameter and $H\approx 2.4\times 10^{-18}\,\mbox{s}$ is the Hubble parameter \cite{obs}.
These data determine $\Lambda\approx 1.4\times 10^{-52}\,\mbox{m}^{-2}$, which also gives $b_c\approx 5.6\times 10^{25}\,\mbox{m}$ and $m_c\approx 3.8\times 10^{52}\,\mbox{kg}\approx 1.9\times 10^{22}\,M_\odot$.
The critical black-hole mass is thus estimated to be
\begin{equation}
M_c\approx 35\,M_\odot.
\label{mc}
\end{equation}
Such a value is larger (but not too much) than the mass of the most massive observed stellar black hole (in the binary IC 10 X-1), which is about 24-33 $M_\odot$ \cite{maxBH}.
Additional processes of mass generation in the Universe can, however, effectively decrease $M_c$.
For example, adding to the energy-momentum tensor a term that is proportional to the symmetrized covariant derivative of the four-velocity of matter will increase $m$ \cite{Lo,cre}.
As a result, a universe inside a typical stellar black hole can build up its mass during oscillations and, after a finite number of cycles, expand to infinity.
The mass of such a universe would be larger than that in (\ref{masuni}).

Another possibility to form a universe expanding to infinity from the interior of a stellar black hole comes from stellar binaries or multiple star systems containing a black hole.
Since the event horizon's formation and the subsequent formation of a universe inside occur after an infinite amount of time in the frame of reference of external observers, they occur after such systems have merged.
The matter in a black hole that forms from merging ultradense stars will also be described by a stiff equation of state.
The interiors of forming intermediate-mass black holes should be composed of stiff matter too.
If the total mass of such a system merging into one black hole exceeds $M_c$ in (\ref{mc}) (several massive stars would suffice) then the universe inside this black hole will expand to infinity.

Astronomical observations \cite{obs} determine the scale factor of the present Universe to be $a_\textrm{p}\approx 2.9\times 10^{27}\,\mbox{m}$ \cite{infl}.
The mass of the Universe is approximately $m\approx 4\pi \Omega_\textrm{M}\rho_c a_\textrm{p}^3/3$, where $\Omega_\textrm{M}\approx 0.26$ is the matter density parameter and $\rho_c=3H^2/(8\pi G)$ is the critical density of the Universe.
Thus we find $m\approx 2.7\times 10^{56}\,\mbox{kg}\approx 1.4\times 10^{26}\,M_\odot$.
Equaling this mass to $m_\textrm{max}$ in (\ref{mmax}) gives
\begin{equation}
M\approx 3\times 10^3\,M_\odot>M_c,
\end{equation}
which is within the range of intermediate-mass black holes.
Therefore, our Universe may have originated from the interior of an intermediate-mass black hole.

Finally, we briefly discuss cosmological perturbations \cite{general} in the Universe born in a black hole that has formed from collapsing spin-fluid matter with a stiff equation of state.
Scalar perturbations originating from quantum fluctuations in the initial vacuum state that existed in the contracting phase of the Universe and was associated with the hydrodynamical fluctuations have been considered in \cite{stiff}.
The expanding phase of the early Universe in that bouncing model was described by a scale factor whose dependence on the conformal time $\eta$, $a(\eta)=a(0)(1+\eta^2/\eta_0^2)^{1/2}$, is the same as in \cite{infl}.
The perturbations produced in \cite{stiff} are nearly scale-invariant if the Universe had a slowly contracting phase, $a(\eta)\propto(-\eta)^k$, where $0<k\ll 1$ (stiff matter gives $k=1/2$).
Thus, initial quantum fluctuations in a universe formed in a black hole cannot produce a scale-invariant power spectrum of perturbations.

Thermal fluctuations in a black hole, however, can lead to such a spectrum.
Such fluctuations in a gas of point particles with a barotropic equation of state $p=w_r\epsilon$ produce a scale-invariant power spectrum if the equation of state of the background, $p=w\epsilon$, satisfies a condition $w_r=(w-1)/4$ \cite{thermal}.
If the background has a stiff equation of state, $w=1$, then thermal fluctuations with a nonrelativistic equation of state $w_r=0$ in a collapsing black hole can generate scale-invariant perturbations.
If the nonrelativisitic-matter component causing fluctuations is produced during the contracting phase at a sufficient rate then it will not dilute out due to scaling.
Moreover, at extremely high densities, fermionic matter may be dominated by very massive particles.
We also mentioned that fermions at such densities acquire additional effective masses due to the cubic term in the Hehl-Datta equation, supporting the nonrelativisitic-matter component of the Universe.
The perturbations resulting from $w=1$ and $w_r=0$ could thus propagate through the bounce as a dominant mode.
Therefore, the observed scale-invariant power spectrum of cosmological perturbations is, in principle, consistent with a scenario in which our Universe was born in a black hole.


\end{document}